\documentclass[10pt,conference,letterpaper]{IEEEtran}
\IEEEoverridecommandlockouts

\usepackage{times,amsmath,epsfig,amssymb,graphicx,amsfonts,amsthm}
\usepackage{subfigure}
\usepackage{empheq}
\usepackage{graphicx}
\usepackage{psfrag}
\usepackage{fge}
\usepackage{amsmath}
\usepackage{amsthm}
\usepackage{color}
\usepackage{amsfonts}   
\usepackage{amssymb}    
\usepackage{mathrsfs}
\usepackage{setspace}
\usepackage{dblfloatfix}
\usepackage{tikz}
\usepackage{tkz-tab}
\usepackage{algorithm}
\usepackage{algcompatible}
\usepackage{lipsum}
\usepackage{color}
\usepackage{mathrsfs}
\usepackage{epstopdf}
\usepackage{enumerate}

\floatname{algorithm}{Algorithm}

\newtheorem{mydef}{Definition}
\newtheorem{myassump}{Assumption}
\newtheorem{mytheorem}{Theorem}

\newtheorem{myproposition}{Proposition}

\newcounter{ale}

\newenvironment{liste}{\begin{itemize}}{\end{itemize}}
\newcommand{\aliste}{\begin{liste} \setcounter{ale}{1}}
\newcommand{\zliste}{\end{liste}}

\title{{\LARGE {\bf Minimal Actuator Placement with Optimal Control Constraints}}}
\author{V.~Tzoumas, M.~A.~Rahimian, G.~J.~Pappas, A.~Jadbabaie{$^{\star}$}
\thanks{$^{\star}$All authors are with the Department of Electrical and Systems Engineering, University of Pennsylvania, Philadelphia, PA 19104-6228 USA (email: {\fontsize{8}{8}\selectfont\ttfamily\upshape \{vtzoumas, mohar, pappasg, jadbabai\}@seas.upenn.edu}).}
\thanks{This work was supported in part by TerraSwarm, one of six centers of STARnet, a Semiconductor Research Corporation program sponsored by MARCO and DARPA, and in part by AFOSR Complex Networks Program.}
}

\begin{document}
\maketitle

\begin{abstract}
We introduce the problem of minimal actuator placement in a linear control system so that a bound on the minimum control effort for a given state transfer is satisfied while controllability is ensured.  We first show that this is an NP-hard problem following the recent work of Olshevsky~\cite{2013arXiv1304.3071O}.  Next, we prove that this problem has a supermodular structure.  Afterwards, we provide an efficient algorithm that approximates up to a multiplicative factor of $O(\log n)$, where $n$ is the size of the multi-agent network, any optimal actuator set that meets the specified energy criterion.  Moreover, we show that this is the best approximation factor one can achieve in polynomial-time for the worst case.  Finally, we test this algorithm over large Erd\H{o}s-R\'{e}nyi random networks to further demonstrate its efficiency.
\end{abstract}


\section{Introduction}\label{sec:Intro}

During the past decade, control scientists have developed various tools for the regulation of large-scale systems, with the notable examples of~\cite{orosz2010controlling} for the control of biological systems,~\cite{ching2012distributed} for the regulation of brain and neural networks,~\cite{2013arXiv1309.6270P} for network protection against spreading processes, and~\cite{Chen-2012-DR-Springer} for load management in smart grid.  On the other hand, the enormous size of these systems and the need for cost-effective control make the identification of a small fraction of their nodes to steer them around the state space a central problem within the control community~\cite{ 2013arXiv1304.3071O,citeulike:13239948, ramos2014np, 2014arXiv1404.7665S, bullo2014,2014arXiv1409.3289T}.  

This is a combinatorial task of formidable complexity; as it is shown in~\cite{2013arXiv1304.3071O}, identifying a small set of actuator nodes so that the resultant system is controllable alone is NP-hard. Nonetheless, a controllable system may be practically uncontrollable if the required input energy for the desired state transfers is forbidding, as when the controllability matrix is close to singularity~\cite{Chen:1998:LST:521603}.  Therefore, by choosing input nodes to ensure controllability alone, one may not achieve a cost-effective control for the involved state transfers.
{In this paper, we aim to address this important requirement, by introducing a best-approximation polynomial-time algorithm to actuate a small fraction of a
system's states so that controllability is ensured and a specified control energy performance is guaranteed.}

In particular, we consider the selection of a minimal number of actuators such that a bound on the minimum control effort for a given transfer is satisfied while controllability is ensured. Finding the appropriate choice of such a subset of nodes is a challenging task, since the search for a subset satisfying certain criteria constitutes a combinatorial optimization problem that can be computationally intensive.  Indeed, it is shown in~\cite{2013arXiv1304.3071O} that identifying the minimum number of actuators for inducing controllability alone is NP-hard.  Therefore, we extend this computationally hard problem by imposing an energy constraint on the choice of the actuator set, and we solve it with an efficient approximation algorithm.

Specifically, we first generalize the involved energy objective to an $\epsilon$-close one, which remains well-defined even when the controllability matrix is non-invertible.  Then, we make use of this metric and we relax the controllability constraint of the original problem.  Notwithstanding, we show that for certain values of $\epsilon$ all solutions of this auxiliary program still render the system controllable.  This fact, along with a supermodularity property of the generalized objective that we establish, leads to a polynomial-time algorithm that approximates up to a multiplicative factor of $O(\log n)$ any optimal actuator set that meets the specified energy bound, when the latter lies in a certain range with respect to $n$.
Moreover, we show that this is the best approximation factor one can achieve in polynomial-time for the worst case.  

Hence, with this algorithm we aim to address the open problem of actuator placement with energy performance guarantees~\cite{2013arXiv1304.3071O, 2014arXiv1404.7665S,bullo2014,PhysRevLett.108.218703,PhysRevLett.110.208701}.  
To the best of our knowledge, we are the first to study the selection of a minimal number of actuators so that a bound on the minimum control effort for a given transfer is satisfied.
Our results are also applicable to the case of average control energy metrics~\cite{Muller1972237} and can be extended to the cardinality-constrained actuator placement for minimum control
effort, where the optimal actuator set is selected so that these
metrics are minimized, while its cardinality is upper bounded by a given value.  These and other relevant extensions are explored in the companion manuscript~\cite{2014arXiv1409.3289T}.

The remainder of this paper is organized as follows. The formulation and model for the actuator selection problem are set forth in Section \ref{sec:Prelim}.  In Section~\ref{sec:Min_N} we discuss our main results, including the intractability of this problem, as well as the supermodularity of the involved control energy objective.  Then, we provide an efficient approximation algorithm for its solution.  Finally, in Section~\ref{sec:examples} we illustrate our analytical findings on an integrator chain network and we test their performance over large Erd\H{o}s-R\'{e}nyi random networks. Section~\ref{sec:conc} concludes the paper. All proofs can be found in the Appendix.

\section{Problem Formulation} \label{sec:Prelim}


\paragraph*{Notation}  Denote the set of natural numbers $\{1,2,\ldots\}$ as $\mathbb{N}$, the set of real numbers as  $\mathbb{R}$,  and let $[n]\equiv \{1, 2, \ldots, n\}$ for all $n \in \mathbb{N}$.  Also, given a set $\mathcal{X}$, denote $|\mathcal{X}|$ as its cardinality.  Matrices are represented by capital letters and vectors by lower-case letters. For a matrix ${A}$, ${A}^{T}$ is its transpose, and $A_{ij}$ is its element located at the $i-$th row and $j-$th column. For a symmetric matrix ${A}$, ${A}={A}^{T}$; and if ${A}$ is positive semi-definite, or positive definite, we write ${A} \succeq {0}$ and ${A}\succ {0}$, respectively.  Moreover, for $i \in [n]$, let ${I}^{(i)}$ be an $n \times n$ matrix with a single non-zero element: $I_{ii}=1$, while $I_{jk}=0$, for $j$, $k\neq i$, and denote the identity matrix by ${I}$, where its dimension is inferred from the context. Additionally, for $\delta \in \mathbb{R}^n$, ${diag}(\delta)$ denotes an $n \times n$ diagonal matrix such that ${diag}(\delta)_{ii}=\delta_i$, for all $i \in [n]$.

\subsection{Actuator Placement Model}

Consider a linear system of $n$ states, $x_1, x_2,\ldots,x_n$, whose evolution is described by
\begin{align}
\dot{{x}}(t) = {A}{{x}}(t) + {B}{{u}}(t), t > t_0,
\label{eq:dynamics}
\end{align} where  $t_0 \in \mathbb{R}$ is fixed, ${x}\equiv \{x_1,x_2,\ldots,x_n\}$, $\dot{{x}}(t)\equiv d{x}/dt$, while ${u}$ is the corresponding input vector.  The matrices ${A}$ and ${B}$ are of appropriate dimension.  Without loss of generality, we also refer to~\eqref{eq:dynamics} as a network of $n$ agents, $1, 2,\ldots, n$, which we associate with the states $x_1, x_2,\ldots, x_n$, respectively.  Moreover, we denote their collection as $\mathcal{V}\equiv[n]$. Henceforth,  the interaction matrix ${A}$ is fixed, while a special structure is assumed for the input matrix ${B}$. 

\begin{myassump}\label{assump:Diag_B}
${B}={diag}(\delta)$, where $\delta\in\{0,1\}^{n}$.
\end{myassump}

Each choice of the binary vector $\delta$ in Assumption~\ref{assump:Diag_B} signifies a particular selection of agents as actuators.  Hence, if $\delta_i=1$, state $i$ may receive an input, while if $\delta_i=0$, receives none.  We collect the above and others into the next definition.

\begin{mydef}[{Actuator Set, Actuator}]
Given $\delta \in \{0,1\}^{n}$ and ${B} = {diag}(\delta)$, let $\Delta \subseteq \mathcal{V}$ be such that $\forall i\in \Delta$, $\delta_i=1$, while $\forall i\notin \Delta$, $\delta_i=0$; then, $\Delta$ is called an \emph{actuator set} and any agent $i \in \Delta$ is called an \emph{actuator}.
\end{mydef}

\subsection{Controllability and the Minimum Energy Transfer Problem}

We consider the notion of controllability and relate it to the problem of selecting a minimum number of actuators for the satisfaction of a control energy constraint.

Recall that~\eqref{eq:dynamics} is controllable if for any finite $t_1>t_0$ and any initial state ${x}_0\equiv {x}(t_0)$, the system can be steered to any other state ${x}_1\equiv {x}(t_1)$, by some input ${u}(t)$ defined over $[t_0, t_1]$.  Moreover, for general matrices ${A}$ and ${B}$, the controllability condition is equivalent to the matrix 
\begin{align}
{\Gamma}(t_0,t_1)\equiv \int_{t_0}^{t_1} \mathrm{e}^{{A}(t-t_0)} {B}{B}^{T} \mathrm{e}^{{A}^{T}(t-t_0)}\,\mathrm{d}{t},\label{eq:general_gramian}
\end{align} being positive definite for any $t_1>t_0$~\cite{Chen:1998:LST:521603}.  Therefore, we refer to ${\Gamma}(t_0,t_1)$ as the \textit{controllability matrix} of~\eqref{eq:dynamics}.

The controllability of a linear system is of great interest, because it is related to the solution of the following minimum energy transfer problem

\begin{equation}\label{pr:min_energy_transfer}
\begin{aligned}
 \underset{{u}(\cdot)}{\text{minimize}} & \; \;  \; 
 \int_{t_0}^{t_1} {u}(t)^T{u}(t)\,\mathrm{d}{t}\\
\text{subject to} \\
& \dot{{x}}(t) = {A}{{x}}(t) + {B}{{u}}(t), t_0 <t \leq t_1,\\
& {x}(t_0)= {x_0}, {x}(t_1)={x_1},
\end{aligned}
\end{equation} where ${A}$ and ${B}$ are any matrices of appropriate dimension.  In particular,  if~\eqref{eq:dynamics} is controllable for the given ${A}$ and ${B}$, the resulting minimum control energy is given by
\begin{align}
({x}_1 - \mathrm{e}^{{A}\tau}{x}_0)^{T}{\Gamma}(t_0,t_1)^{-1}({x}_1 - \mathrm{e}^{{A}\tau}{x}_0),\label{exact_energy}
\end{align}
where $\tau=t_1-t_0$~\cite{Muller1972237}. 
Therefore, if ${x}_1 - \mathrm{e}^{{A}\tau}{x}_0$ is spanned by the eigenvectors of ${\Gamma}(t_0,t_1)$ corresponding to its smallest eigenvalues, the minimum control effort~\eqref{exact_energy} may be forbiddingly high~\cite{Chen:1998:LST:521603}.  Hence, when we choose the actuators of a network so that controllability is {ensured} and an input energy constraint for a specified state transfer is satisfied, we should take into account their effect on ${\Gamma}(t_0,t_1)^{-1}$.

Moreover, controllability is an indispensable property for any linear system, while in many cases is viewed as a structural attribute of the involved system~\cite{1100557} that holds true even by any single input nodes, as in large-scale neural networks~\cite{citeulike:13239948}.  This motivates further the setting of this paper, where the actuators are chosen so that a bound on the minimum control effort for a given transfer is satisfied and overall controllability is respected.


Per Assumption~\ref{assump:Diag_B} some further properties for the controllability matrix are due.  First, given an actuator set $\Delta$, associated with some $\delta$, let ${\Gamma}_\Delta \equiv {\Gamma}(t_0,t_1)$; then, 
\begin{align}
{\Gamma}_{\Delta} = \sum_{i=1}^n \delta_i {\Gamma}_i,
\label{eq:gramianTOdelta}
\end{align}where for any $i \in [n]$, ${\Gamma}_i = \int_{t_0}^{t_1} \mathrm{e}^{{A}t} {I}^{(i)} \mathrm{e}^{{A}^{T} t}\,\mathrm{d}{t}$, that is, each ${\Gamma}_i$ is a constant positive semi-definite matrix determined by ${A}$, $t_0$ and $t_1$.  To see why~\eqref{eq:gramianTOdelta} holds true, observe that ${B} = {diag}(\delta)$ implies ${B} = {B} {B}^{T} =  \sum_{i = 1}^{n} \delta_i {I}^{(i)}$, and \eqref{eq:gramianTOdelta} follows upon replacing this in~\eqref{eq:general_gramian}. 
Furthermore, note that \eqref{eq:gramianTOdelta} together with the fact that ${\Gamma}_i \succeq {0}$, for any $i \in [n]$ gives  ${\Gamma}_{\Delta_1}\preceq {\Gamma}_{\Delta_2} $ whenever $ {\Delta_1}\subseteq{\Delta_2}$.

\subsection{Actuator Placement Problem}\label{subsec:leader_pr}

We consider the problem of actuating a small number of system's~\eqref{eq:dynamics} states so that the minimum control energy for a given transfer meets some specified criterion and controllability is ensured. The challenge is in doing so using as few actuators as possible.  This is an important improvement over the existing literature where the goal of actuator placement problems have either been to ensure just controllability~\cite{2013arXiv1304.3071O} or the weaker property of structural controllability~\cite{jafari2011leader,Commault20133322}.  
Other relevant results consider the task of leader-selection~\cite{clark2014_2,clark2014_1}, where the leaders, i.e. actuated agents, are chosen so as to minimize an appropriate mean-square convergence error of the remaining agents.  
Our work also departs from a set of works that study average energy metrics, such as the minimum eigenvalue of the controllability Gramian or the trace of its inverse~\cite{2014arXiv1404.7665S,bullo2014,PhysRevLett.108.218703}.  Instead, here we consider an exact energy objective and require it to satisfy a particular upper bound.

Let $\mathcal{C}_r \equiv \{\Delta \subseteq \mathcal{V}: |\Delta| \leq r, {\Gamma}_\Delta \succ 0\}$ be the actuator sets of cardinality at most $r$ that render~\eqref{eq:dynamics} controllable.   Then, for any $\Delta \subseteq \mathcal{V}$, we write $\Delta \in \mathcal{C}_{|\Delta|}$ to denote that $\Delta$ achieves controllability.   Furthermore, we set
\[
{v}\equiv ({x}_1 - \mathrm{e}^{{A}\tau}{x}_0)/\|{x}_1 - \mathrm{e}^{{A}\tau}{x}_0\|_2.
\]

We consider the problem

\begin{equation}\tag{I}\label{pr:min_set}
\begin{aligned}
 \underset{\Delta \subseteq \mathcal{V}}{\text{minimize}} & \; \;  \; 
 |\Delta|\\
\text{subject to} \\
&  \Delta \in \mathcal{C}_{|\Delta|},\\ 
&{v}^{T}{\Gamma}_\Delta^{-1}{v} \leq E,
\end{aligned}
\end{equation}  for some positive constant $E$. This problem is a generalized version of the minimal controllability problem considered in~\cite{2013arXiv1304.3071O}, so that its solution not only ensures controllability, but also provides a guarantee in terms of the minimum input energy required for the normalized transfer from ${x}_0$ to ${x}_1$; indeed, for $E\rightarrow\infty$, we recover the problem of~\cite{2013arXiv1304.3071O}.  

For some extra properties of~\eqref{pr:min_set}, note that for any $\Delta \in \mathcal{C}_{|\Delta|}$, $0 \prec {\Gamma}_\Delta \preceq {\Gamma}_\mathcal{V}$, i.e. ${v}^{T}{\Gamma}_\mathcal{V}^{-1}{v}\leq {v}^{T}{\Gamma}_\Delta^{-1}{v}$~\cite{bernstein2009matrix}.  Hence,~\eqref{pr:min_set} is feasible for any $E$ such that
\begin{align}
{v}^{T}{\Gamma}_\mathcal{V}^{-1}{v}\leq E.\label{lo_bound_E}
\end{align} 
Observe that this lower bound depends only on $A$ and $v$, i.e. also on $n$, as well as on $t_0$ and $t_1$.

Moreover,~\eqref{pr:min_set} is NP-hard, since it looks for a minimal solution and so it asks if $\mathcal{C}_r \neq \emptyset$ for any $r< n$~\cite{2013arXiv1304.3071O}.   Thus, we need to identify an efficient approximation algorithm for its solution, which is the subject of the next section.

\section{Minimal Actuator Sets with Constrained Minimum Energy Performance}\label{sec:Min_N}

{We present} an efficient polynomial-time approximation algorithm for~\eqref{pr:min_set}.  To this end, we first generalize the involved energy objective to an $\epsilon$-close one, that remains well-defined even when the controllability matrix is non-invertible.  Next, we relax~\eqref{pr:min_set} {by introducing a program that makes use of this objective and ignores controllability constraint of \eqref{pr:min_set}.  Nonetheless, we show that for certain values of $\epsilon$ all solutions of this auxiliary program still render the system controllable.  This fact, along with the supermodularity property of the generalized objective that we establish, leads to our proposed approximation algorithm.  The discussion of its efficiency ends the analysis of~\eqref{pr:min_set}.

\subsection{An $\epsilon$-close Auxiliary Problem}\label{subsubsec:Min_N_aux}

Consider the following approximation to Problem~\eqref{pr:min_set}
\begin{equation}\tag{I$'$}\label{pr:min_set_approx}
\begin{aligned}
 \underset{\Delta \subseteq \mathcal{V}}{\text{minimize}} & \; \;  \; 
 |\Delta|\\
\text{subject to} \\
&  \phi(\Delta) \leq E,
\end{aligned}
\end{equation} 
where $
\phi(\Delta)\equiv {v}^{T}({\Gamma}_\Delta+\epsilon{I})^{-1}{v}+ \epsilon\sum_{i=1}^{n-1}\bar{{v}}_i^{T}({\Gamma}_\Delta+\epsilon^2{I})^{-1}\bar{{v}}_i$,  for any $\Delta \subseteq \mathcal{V}$, while $\bar{{v}}_1,\bar{{v}}_2,\ldots,\bar{{v}}_{n-1}$ are an orthonormal basis for the null space of ${v}$, and $\epsilon$ is fixed such that $0< \epsilon \leq 1/E$, given $E$.  Observe that the controllability constraint is now ignored, while the energy objective is well-defined for any actuator set $\Delta$ including the empty set, since the  invertibility of ${\Gamma}_\Delta+\epsilon{I}$ and ${\Gamma}_\Delta+\epsilon^2{I}$ is always guaranteed for $\epsilon > 0$.  

The $\epsilon$-closeness is evident, since for any
$\Delta \in \mathcal{C}_{|\Delta|}$, $ \phi(\Delta)\rightarrow {v}^{T}{\Gamma}_\Delta^{-1}{v}$ as $\epsilon\rightarrow 0$.  Notice that we can take $\epsilon \rightarrow 0$, since we assume any positive $\epsilon \leq 1/E$.  

\subsection{Approximation Algorithm for Problem~\eqref{pr:min_set_approx}} \label{subsubsec:Min_N_alg}

We first prove that all solutions of~\eqref{pr:min_set_approx} for $0<\epsilon \leq 1/E$, render the system controllable, notwithstanding that no controllability constraint is imposed by this program on the choice of the actuator sets.  Moreover, we show that the involved $\epsilon$-close energy objective is supermodular, and then we present our approximation algorithm, followed by a discussion of its efficiency, which ends this subsection.

\begin{myproposition}\label{prop:suf_contr}
Fix $\omega>0$.  Then, $\forall \epsilon$, $0< \epsilon\leq 1/\omega$, if $\forall\Delta \subseteq \mathcal{V}$, $\phi(\Delta) \leq \omega$, then $\Delta \in \mathcal{C}_{|\Delta|}$.
\end{myproposition}

Note that $\omega$ is chosen independently of the parameters of system~\eqref{eq:dynamics}.  Therefore,  the absence of the controllability constraint at Problem~\eqref{pr:min_set_approx} for $0<\epsilon \leq 1/E$ is fictitious; nonetheless, it obviates the necessity of considering only those actuator sets that render the system controllable. 

The next lemma is also essential and suggest an efficient approximation algorithm for solving~\eqref{pr:min_set_approx}.

\begin{myproposition}[Supermodularity]\label{prop:subm}
The function ${v}^{T}({\Gamma}_\Delta+\epsilon{I})^{-1}{v}+ \epsilon\sum_{i=1}^{n-1}\bar{{v}}_i^{T}({\Gamma}_\Delta+\epsilon^2{I})^{-1}\bar{{v}}_i: \Delta \subseteq \mathcal{V} \mapsto \mathbb{R}$ is supermodular with respect to the choice of $\Delta$.
\end{myproposition}

Inspired by the literature on set-covering problems subject to submodular constraints,~\cite{Nemhauser:1988:ICO:42805, citeulike:416650,krause2012submodular}, we have the following efficient approximation algorithm for Problem~\eqref{pr:min_set_approx}, and, as we illustrate by the end of this section, for Problem~\eqref{pr:min_set} as well. We note that a corollary of the above proposition is that ${v}^{T}{\Gamma}_{(\cdot)}^{-1}{v}$ is supermodular as well, but over the sets $\Delta\subseteq \mathcal{V}$ that render~\eqref{eq:dynamics} controllable.

\begin{algorithm}
\caption{Approximation Algorithm for the Problem~\eqref{pr:min_set_approx}.}\label{alg:minimal-leaders}
\begin{algorithmic}
\REQUIRE Upper bound $E$, approximation parameter $\epsilon \leq 1/E$, matrices ${\Gamma}_1, {\Gamma}_2, \ldots, {\Gamma}_n$, vector ${v}$.
\ENSURE Actuator set $\Delta$.
\STATE $\Delta\leftarrow\emptyset$
\WHILE {$\phi(\Delta) > E $} \STATE{ 	
     $a_i \in \text{argmax}_{a \in \mathcal{V}\setminus \Delta}\{
	\phi(\Delta)-\phi(\Delta\cup\{a\}) 
	\}$\\
	\quad \mbox{}  $\Delta \leftarrow \Delta \cup \{a_i\}$	
	}
\ENDWHILE
\end{algorithmic} 
\end{algorithm}

For the efficiency of Algorithm~\ref{alg:minimal-leaders} the following is true. 

\begin{mytheorem}[A Submodular Set Coverage Optimization]\label{th:minimal}
Denote as $l^\star$ the cardinality of a solution to Problem~\eqref{pr:min_set_approx} and as $\Delta$ the selected set by Algorithm~\ref{alg:minimal-leaders}.  Then,
\begin{align}
&\Delta  \in \mathcal{C}_{|\Delta|},\label{explain:th:minimal2}\\
&\phi(\Delta) \leq E,\label{explain:th:minima3}\\
&\frac{|\Delta|}{l^\star}\leq 1+\log \frac{n\epsilon^{-1}-\phi(\mathcal{V})}{E-\phi(\mathcal{V})}\equiv F, \label{explain:th:minimal1}\\
&F=O(\log n + \log \epsilon^{-1}+\log \frac{1}{E-\phi(\mathcal{V})}).\label{explain:approx_error0}
\end{align}
\end{mytheorem}

Therefore, the polynomial-time Algorithm~\ref{alg:minimal-leaders} returns a set of actuators that meets the corresponding control energy bound of Problem~\eqref{pr:min_set_approx}, while it renders system~\eqref{eq:dynamics} controllable.  Moreover, the cardinality of this set is up to a multiplicative factor of $F$ from the minimum cardinality actuator sets that meet the same control energy bound.  In Section~\ref{sebsec:Quality} we elaborate further on the dependence of this multiplicative factor on $n$, $\epsilon$ and $E$, using~\eqref{explain:approx_error0}, while in Section~\ref{subsec:ApproximationAlgorithm} we finalize our treatment of Problem~\eqref{pr:min_set} by employing Algorithm~\ref{alg:minimal-leaders} to approximate its solutions.

\subsection{Quality of Approximation of Algorithm~\ref{alg:minimal-leaders} for Problem~\eqref{pr:min_set_approx}}\label{sebsec:Quality}

The result in~\eqref{explain:approx_error0} was expected from a design perspective: Increasing the network size $n$ or improving the accuracy by decreasing $\epsilon$, as well as demanding a better energy guarantee by decreasing $E$, should all push the cardinality of the selected actuator set upwards. Also, note that $\log \epsilon^{-1}$ is the design cost for circumventing the difficulty to satisfy controllability constraint of Problem~\eqref{pr:min_set} directly~\cite{2013arXiv1304.3071O}. 

Furthermore, per~\eqref{explain:approx_error0} and with $E-\phi(\mathcal{V})$ and $\epsilon$ both fixed,  the cardinality of the actuator set that Algorithm~\ref{alg:minimal-leaders} returns is up to a multiplicative factor of $O(\log n)$ from the minimum cardinality actuator sets that meet the same performance criterion.  We note that this is the best achievable bound in polynomial-time for the set covering problem in the worst case~\cite{Feige:1998:TLN:285055.285059}, while~\eqref{pr:min_set_approx} is a generalization of it (cf.~\cite{2013arXiv1304.3071O}). 

\subsection{Approximation Algorithm for Problem~\eqref{pr:min_set}}\label{subsec:ApproximationAlgorithm}

We present an efficient approximation algorithm for Problem~\eqref{pr:min_set} that is based on Algorithm~\ref{alg:minimal-leaders}.  To this end, let $\Delta$ be the actuator set returned by Algorithm~\ref{alg:minimal-leaders}, i.e. $\Delta \in \mathcal{C}_{|\Delta|}$ and $\phi(\Delta)\leq E$.   Moreover, denote as $\lambda_1$, $\lambda_2$, $\ldots$, $\lambda_n$ and ${q}_1$, ${q}_2$, $\ldots$, ${q}_n$ the eigenvalues and the corresponding orthonormal eigenvectors of ${\Gamma}_{\Delta}$, respectively.  Additionally,  let $\lambda_m\equiv\text{min}_{i\in [n]}\lambda_i$ and ${q}_M\equiv \text{argmax}_{{q_i}, i\in[n]} {v}^T{q_i}$.  Finally, consider a positive $\epsilon$ such that $n\epsilon({v}^T{q}_M)^2/\lambda_m^2\leq cE$, for some $c>0$. Then,
\begin{align}
\phi(\Delta)>{v}^{T}({\Gamma}_\Delta+\epsilon{I})^{-1}{v}&=\sum_{j=1}^{n}\frac{({v}^T{q_j})^2}{\lambda_j+\epsilon} \label{eq:aux_1}\\
&\geq{v}^{T}{\Gamma}_\Delta^{-1}{v}-\frac{n\epsilon({v}^T{q}_M)^2}{\lambda_m^2}\label{eq:aux_2}\\
&\geq {v}^{T}{\Gamma}_\Delta^{-1}{v}-cE, \label{eq:aux_3}
\end{align} where we derived~\eqref{eq:aux_2} from~\eqref{eq:aux_1} using the fact that for any $x\geq 0$, $1/(1+x)\geq 1-x$, while the rest follow from the definition of $\lambda_m$ and ${q}_M$, as well as the assumption $n\epsilon({v}^T{q}_M)^2/\lambda_m^2\leq cE$.  Moreover, it is also true that $\phi(\Delta)\leq E$ by the definition of $\Delta$, and therefore from~\eqref{eq:aux_3} we get
\begin{align}
{v}^{T}{\Gamma}_\Delta^{-1}{v}\leq (1+c)E. \label{eq:approx_error}
\end{align} Hence, we refer to $c$ as \textit{approximation error}.  

On the other hand, $\lambda_m$ and ${q}_M$ are not in general known in advance.  Hence, we need to search for a sufficiently small value of $\epsilon$ so that~\eqref{eq:approx_error} holds.  One way to achieve this, since $\epsilon$ is lower and upper bounded by $0$ and $1/E$, respectively, is to perform a binary search.  We implement this procedure in Algorithm~\ref{alg:minimal-leaders_final}, where we denote as  $[\text{Algorithm}~\ref{alg:minimal-leaders}](E,\epsilon)$ the set that Algorithm~\ref{alg:minimal-leaders} returns, for given $E$ and $\epsilon$.  

\begin{algorithm}
\caption{Approximation Algorithm for the Problem~\eqref{pr:min_set}.}\label{alg:minimal-leaders_final}
\begin{algorithmic}
\REQUIRE Upper bound $E$, approximation error $c$, bisection's accuracy level $a$, matrices ${\Gamma}_1, {\Gamma}_2, \ldots, {\Gamma}_n$, vector ${v}$.
\ENSURE Actuator set $\Delta$.
\STATE $l\leftarrow 0$, $u\leftarrow 1/E$, $\epsilon\leftarrow(l+u)/2$
\WHILE {$u-l>a$}\\	
    $\Delta \leftarrow [\text{Algorithm}~\ref{alg:minimal-leaders}](E,\epsilon)$
	\IF {${v}^{T}{\Gamma}_\Delta^{-1}{v}- {v}^{T}({\Gamma}_\Delta+\epsilon{I})^{-1}{v}> cE$}\STATE{$u\leftarrow \epsilon$} \ELSE \STATE{$l\leftarrow \epsilon$} \ENDIF\\
	$\epsilon\leftarrow (l+u)/2$   
\ENDWHILE
\IF {${v}^{T}{\Gamma}_\Delta^{-1}{v}- {v}^{T}({\Gamma}_\Delta+\epsilon{I})^{-1}{v}> cE$} \STATE{$u\leftarrow \epsilon$}, $\epsilon\leftarrow (l+u)/2$
\ENDIF\\
$\Delta \leftarrow [\text{Algorithm}~\ref{alg:minimal-leaders}](E,\epsilon)$\label{exit_step}
\end{algorithmic} 
\end{algorithm}

Note that in the worst case, when we first enter the \texttt{while} loop, the \texttt{if} condition is not satisfied and as a result, $\epsilon$ is set to a lower value.  This process continues until the \texttt{if} condition is satisfied for the first time, from which point and on, the algorithm converges, up to the accuracy level $a$, to the largest value $\bar{\epsilon}$ of $\epsilon$ such that ${v}^{T}{\Gamma}_\Delta^{-1}{v}- {v}^{T}({\Gamma}_\Delta+\epsilon{I})^{-1}{v}\leq cE$; specifically, $|\epsilon-\bar{\epsilon}| \leq a/2$, due to the mechanics of the bisection.  Then, Algorithm~\ref{alg:minimal-leaders_final} exits the \texttt{while} loop and the last \texttt{if} statement ensures that $\epsilon$ is set below $\bar{\epsilon}$ so that ${v}^{T}{\Gamma}_\Delta^{-1}{v}- {v}^{T}({\Gamma}_\Delta+\epsilon{I})^{-1}{v} \leq cE$. The efficiency of this algorithm for Problem~\eqref{pr:min_set} is summarized below.

\begin{mytheorem}[Approximation Efficiency of Algorithm~\ref{alg:minimal-leaders_final} for Problem~\eqref{pr:min_set}]\label{th:minimal_set_main}
Denote as $l^\star$ the cardinality of a solution to Problem~\eqref{pr:min_set_approx} and as 
$\Delta$ the selected set by Algorithm~\ref{alg:minimal-leaders_final}.  Then, 
\begin{align}
&\Delta  \in \mathcal{C}_{|\Delta|}, \nonumber\\
&{v}^{T}{\Gamma}_\Delta^{-1}{v} \leq (1+c)E, \label{state:1}\\
&\frac{|\Delta|}{l^\star}\leq  F, \label{state:1.5}\\
&F = O(\log n+ \log \frac {1}{c\lambda_mE}+\log \frac{1}{E-\phi(\mathcal{V})}).\label{state:2}
\end{align}
\end{mytheorem}

We remark that as $\epsilon \to 0$, $\phi(\cdot)\to {v}^{T}({\Gamma}_{(\cdot)}+\epsilon{E})^{-1}{v}$.  Therefore, for any solution $\Delta^\circ$ to Problem~\eqref{pr:min_set} and $\epsilon$ small enough,  $|\Delta^\circ| \geq l^\star$; to see this, note that as $\epsilon \to 0$ for any $\Delta^\circ$, $\phi(\Delta^\circ) \to {v}^{T}({\Gamma}_{\Delta^\circ}+\epsilon{I})^{-1}{v} < {v}^{T}{\Gamma}_{\Delta^\circ}^{-1}{v} \leq E$, since also $\epsilon>0$, i.e. any $\Delta^\circ$ is a candidate solution to Problem~\eqref{pr:min_set_approx}, which implies $|\Delta^\circ| \geq l^\star$.  Therefore, for $\epsilon$ small enough we have $|\Delta|/|\Delta^\circ| \leq |\Delta|/l^\star$.  Hence,~\eqref{state:1.5} is written as $|\Delta|/|\Delta^\circ| \leq F$, that is, the worst case bound~\eqref{state:1.5} holds also with respect to the cardinality of any solution to~\eqref{pr:min_set}; and as a result, the best-approximation properties of Algorithm~\ref{alg:minimal-leaders} are inherited by Algorithm~\ref{alg:minimal-leaders_final} as well.

\section{Examples and Discussion}\label{sec:examples}

We test the performance of the proposed algorithm over various systems, starting with an integrator chain in Subsection~\ref{subsec:integratorChain} and following up with Erd\H{o}s-R\'{e}nyi random networks in Subsection~\ref{subsec:randomGraphs}.

\subsection{The Case of an Integrator Chain}\label{subsec:integratorChain}

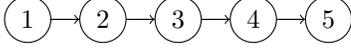
\begin{figure}[t]
\centering
\begin{tikzpicture}
\tikzstyle{every node}=[draw,shape=circle];
\node (v0) at (0:0) {$1$};
\node (v1) at ( 0:1) {$2$};
\node (v2) at ( 0:2) {$3$};
\node (v3) at (0:3) {$4$};
\node (v4) at (0:4) {$5$};

\foreach \from/\to in {v0/v1, v1/v2, v2/v3, v3/v4}
\draw [->] (\from) -- (\to);
\draw
(v0) -- (v1)
(v1) -- (v2)
(v2) -- (v3)
(v3) -- (v4);

\end{tikzpicture}
\caption{A $5$-node integrator chain.}
\label{fig:chain}
\end{figure}

We first illustrate the mechanics and efficiency of Algorithm~\ref{alg:minimal-leaders_final} using the integrator chain in Fig.~\ref{fig:chain}, where we let
\begin{align}
{A} = \left[\begin{array}{ccccc}
-1 & 0 & 0 & 0 & 0\\
1 & -1 & 0 & 0 & 0\\
0 & 1 & -1 & 0 & 0\\
0 & 0 & 1 & -1 & 0\\
0 & 0 & 0 & 1 & -1
\end{array}\right].\nonumber
\end{align}

We first run Algorithm~\ref{alg:minimal-leaders_final} with $E \leftarrow {v}^{T}{\Gamma}_{\{1,5\}}^{-1}{v}$ and $a, c \leftarrow .001$ and examine the transfer from ${x}(0) \leftarrow [0, 0, 0, 0, 0]^T$ to  ${x}(1) \leftarrow [1, 1, 1, 1, 1]^T$.  The algorithm returned the actuator set $\{1,4\}$.  As expected, node $1$ is chosen, and this remains true for any other value of ${x}(1)$, since for a chain network to be controllable, it is necessary and sufficient that node $1$ be actuated.  Additionally, $\{1,4\}$ is the exact best actuator set for achieving this transfer.  This is true because using MATLAB\textsuperscript{\textregistered{}}  we can compute
\begin{align*}
&{v}^{T}{\Gamma}_{\{1\}}^{-1}{v}=5.2486\cdot10^6, {v}^{T}{\Gamma}_{\{1,2\}}^{-1}{v}=2.0860\cdot10^4, \\
&{v}^{T}{\Gamma}_{\{1,3\}}^{-1}{v}=159.9369, {v}^{T}{\Gamma}_{\{1,4\}}^{-1}{v}=159.1712,\\
&{v}^{T}{\Gamma}_{\{1,5\}}^{-1}{v}=2.1086\cdot10^4.
\end{align*} Hence, node $1$ alone does not satisfy the upper bound $E$,  while ${v}^{T}{\Gamma}_{\{1,4\}}^{-1}{v}$ not only satisfies this bound, but it also takes the smallest value among all the actuators sets of cardinality two that induce controllability.  Therefore, $\{1,4\}$ is the best minimal actuator set to achieve the given transfer.

\begin{figure*}[th]
\centering
\hspace*{-30pt}
\includegraphics[width=1.05\textwidth]{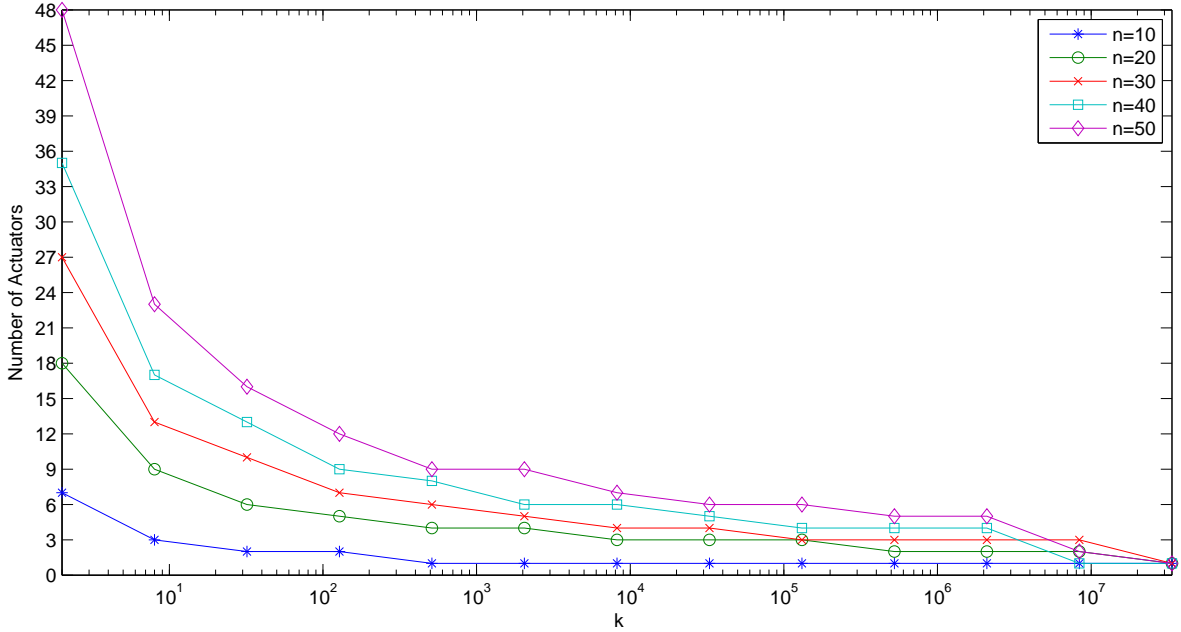}
\caption{Number of selected actuators by Algorithm~\ref{alg:minimal-leaders_final} in Erd\H{o}s-R\'{e}nyi  networks of several sizes $n$ and for varying energy bounds $E$.  For each $n$, the values of $E$ are chosen so that the feasibility constraint~\eqref{lo_bound_E} of Problem~\eqref{pr:min_set} is satisfied: Specifically, for each set of values for $n$ and $k$, Algorithm~\ref{alg:minimal-leaders_final} was executed for $E\leftarrow k{v}^{T}{G(n)}_\mathcal{V}^{-1}{v}$, where $G(n)_\mathcal{V}$ is the controllability Gramian corresponding to the generated network of size $n$ with $B$ set to be the identity matrix.}
\label{fig:randomGraphs}
\end{figure*}

Next, we set ${x}(1) \leftarrow [0, 0, 0, 1, 0]^T$ in Algorithm~\ref{alg:minimal-leaders_final}, which led again to the selection $\{1,4\}$, as one would expect for any transfer that involves only the movement of the fourth node, while controllability is desired.  In other words, even though we chose $E \leftarrow {v}^{T}{\Gamma}_{\{1,5\}}^{-1}{v}$, Algorithm~\ref{alg:minimal-leaders_final} respected this energy bound and with the best possible actuator set for the given transfer, which is $\{1,4\}$, as verified in the following
\begin{align*}
&{v}^{T}{\Gamma}_{\{1\}}^{-1}{v}=1.5425\cdot10^7, {v}^{T}{\Gamma}_{\{1,2\}}^{-1}{v}=5.8675\cdot10^4, \\
&{v}^{T}{\Gamma}_{\{1,3\}}^{-1}{v}=401.7997,{v}^{T}{\Gamma}_{\{1,4\}}^{-1}{v}=6.2889,\\
&{v}^{T}{\Gamma}_{\{1,5\}}^{-1}{v}=2.7445\cdot10^5.
\end{align*}
Moreover, note that although node $1$ is selected as an actuator, in this case its corresponding input signal is zero.  Thus, one may choose not to implement an actuator at this node, at the expense, however, of losing the overall network controllability.  This observation motivates the analysis of~\eqref{pr:min_set} when no controllability constraint is placed on the end actuator set.  

Finally, by setting $E$ large enough in Algorithm~\ref{alg:minimal-leaders_final}, so that any actuator set respects this energy bound, we observe that only node $1$ is selected, as expected for the satisfaction of the controllability constraint.

\subsection{Erd\H{o}s-R\'{e}nyi Random Networks}\label{subsec:randomGraphs}

Erd\H{o}s-R\'{e}nyi random graphs are commonly used to model real-world networked systems~\cite{newman2006structure}.  According to this model, each edge is included in the generated graph with some probability $p$, independently of every other edge.

We implemented this model for varying network sizes $n$, as shown in Fig.~\ref{fig:randomGraphs}, where the directed edge probabilities were set to $p = 2\log(n)/n$, following~\cite{2013arXiv1304.3071O}. 
In particular, we first generated the binary adjacency matrices for each network size so that every edge is present independently  with probability $p$, and then we replaced every non-zero entry with an independent standard normal variable to generate a randomly weighted graph.

To avoid the computational difficulties associated with the integral equation~\eqref{eq:general_gramian} we worked with the controllability Gramian instead, which for a stable system can be efficiently calculated from the Lyapunov equation 
${A}{G} + {G}{A}^{T} = -{B}{B}^{T}$ and is given in closed-form by
\begin{align}
{G} = \int_{t_0}^{\infty} \mathrm{e}^{{A}(t-t_0)} {B} {B}^{T} \mathrm{e}^{{A}^{T}(t-t_0)}\,\mathrm{d}{t}. \nonumber
\end{align} 
Using the controllability Gramian in~\eqref{exact_energy} corresponds to the minimum state transfer energy with no time constraints.
Therefore, we stabilized each random instances of $A$ by subtracting $1.1$ times the real part of their right-most eigenvalue and then we used the MATLAB\textsuperscript{\textregistered{}} function {\fontsize{10}{10}\selectfont\ttfamily\upshape gram} to compute the corresponding controllability Gramians.
 
Next, we set $x_0$ to be the zero vector and $x_1$ the vector of all ones.  We also set $c \leftarrow 0.1$ and $a\leftarrow1$. Finally, for each instance of $n$ we first computed the corresponding lower bound of $E$ so that~\eqref{pr:min_set} is feasible, ${v}^{T}{G}_\mathcal{V}^{-1}{v}$, and then run Algorithm~\ref{alg:minimal-leaders_final} for $E$ equal to $k {v}^{T}{G}_\mathcal{V}^{-1}{v}$, where $k$ ranged from $2$ to $2^{25}$. 

The number of selected actuator nodes by Algorithm~\ref{alg:minimal-leaders_final} for each $n$ with respect to $k$ is shown in Fig.~\ref{fig:randomGraphs}.  We observe that as $k$ increases the number of actuators decreases, as one would expect when the energy bound of~\eqref{pr:min_set} is relaxed.  In addition, we notice that for $k$ large enough, so that~\eqref{pr:min_set} becomes equivalent to the minimal controllability problem of~\cite{2013arXiv1304.3071O}, the number of chosen actuators is one, as it was generally observed in~\cite{2013arXiv1304.3071O} for a similar set of simulations.  

\section{Concluding Remarks}\label{sec:conc}

We introduced the problem of minimal actuator placement in a linear system so that a bound on the minimum control effort for a given state transfer is satisfied while controllability is ensured.  
This problem was shown to be NP-hard and to have a supermodular structure.  Moreover, an efficient algorithm was provided for its solution.  
Finally, the efficiency of this algorithm was illustrated over large Erd\H{o}s-R\'{e}nyi random networks. 
Our future work is focused on investigating the case where no controllability constraint is placed on the end actuator set, as well as, on exploring the effects that the network topology has on this selection.  


\appendices

\section{Proofs of the Main Results}\label{proofs}

\subsection{Proposition~\ref{prop:suf_contr}}

Note that since $\|{v}\|_2=1$ and $\bar{{v}}_1,\bar{{v}}_2,\ldots,\bar{{v}}_{n-1}$ are an orthonormal basis for the null space of ${v}$, for any unit vector ${q}$ of dimension equal to that of ${v}$ it is $({v}^T{q})^2+\sum_{i=1}^{n-1}(\bar{{v}}_i^T{q})^2=1$.  Next, assume that $\Delta \notin \mathcal{C}_{|\Delta|}$ and let $k$ be the corresponding number of non-zero eigenvalues of ${\Gamma}_\Delta$.  Therefore, $k \leq n-1$.  Moreover, denote as $\lambda_1, \lambda_2, \ldots, \lambda_n$ and ${q_1}, {q_2}, \dots, {q_n}$ the eigenvalues and orthonormal eigenvectors of ${\Gamma}_\Delta$.  We get
\begin{align}
\phi(\Delta)=\sum_{j=1}^{k}[\frac{({v}^T{q_j})^2}{\lambda_j+\epsilon}+\sum_{i=1}^{n-1}\frac{\epsilon(\bar{{v}}_i^T{q_j})^2}{\lambda_j+\epsilon^2}]+\frac{n-k}{\epsilon}\geq v.  \nonumber
\end{align} Since $\epsilon \leq 1/v$ and $k \leq n-1$ we have a contradiction. \hfill{}{\scriptsize $\blacksquare$}{\scriptsize \par}

\subsection{Proposition~\ref{prop:subm}} 

We first prove that ${v}^{T}({\Gamma}_\Delta+\epsilon{I})^{-1}{v}$ is supermodular.  With similar steps one can show that  $\bar{{v}}_i^{T}({\Gamma}_\Delta+\epsilon^2{I})^{-1}\bar{{v}}_i$, for any $i\in[n-1]$, also is.  Then, the proof is complete, since the class of supermodular functions is closed under non-negative linear combinations.

Recall that ${v}^{T}({\Gamma}_\Delta+\epsilon{I})^{-1}{v}$ is supermodular if and only if $-{v}^{T}({\Gamma}_\Delta+\epsilon{I})^{-1}{v}$ is submodular, and that a function $h: \mathcal{V}\mapsto \mathbb{R}$ is submodular if and only if for any $a \in \mathcal{V}$ the function $h_a:\mathcal{V}\setminus\{a\}\mapsto \mathbb{R}$, where $h_a(\Delta)\equiv h(\Delta\cup \{a\})-h(\Delta)$, is a non-increasing set function. In other words, if and only if for any $\Delta_1 \subseteq \Delta_2 \subseteq \mathcal{V}\setminus\{a\}$ it holds true that $h_a(\Delta_1)\geq h_a(\Delta_2)$.

In our case, $h_a(\Delta)= -{v}^{T}({\Gamma}_{\Delta\cup \{a\}}+\epsilon{I})^{-1}+{v}^{T}({\Gamma}_\Delta+\epsilon{I})^{-1}{v}$.  Therefore, take any $\Delta_1 \subseteq \Delta_2 \subseteq \mathcal{V}\setminus\{a\}$ and denote accordingly $\mathcal{D}\equiv\Delta_2\setminus\Delta_1$.  Then, we aim to prove
\begin{align*}
&-{v}^{T}({\Gamma}_{\Delta_1\cup \{a\}}+\epsilon{I})^{-1}{v}+{v}^{T}({\Gamma}_{\Delta_1}+\epsilon{I})^{-1}{v}\geq \\
&-{v}^{T}({\Gamma}_{\Delta_1\cup \mathcal{D}\cup \{a\}}+\epsilon{I})^{-1}{v}+{v}^{T}({\Gamma}_{\Delta_1\cup \mathcal{D}}+\epsilon{I})^{-1}{v}.
\end{align*}
To this end and for $z\in [0,1]$, set $f(z)={v}^{T}({\Gamma}_{\Delta_1}+z{\Gamma}_{\mathcal{D}}+{\Gamma}_ a+\epsilon{I})^{-1}{v}$, and $g(z)={v}^{T}({\Gamma}_{\Delta_1}+z{\Gamma}_{\mathcal{D}}+\epsilon{I})^{-1}{v}$. After some manipulations the above inequality can be written as $f(1)-f(0)\geq g(1)-g(0)$. 

To prove this one, it suffices to prove that $df/dz \geq dg/dz$, $\forall z\in (0,1)$.  Denote ${L}_1(z)={\Gamma}_{\Delta_1}+z{\Gamma}_{\mathcal{D}}+{\Gamma}_a+\epsilon{I}$ and ${L}_2(z)={\Gamma}_{\Delta_1}+z{\Gamma}_{\mathcal{D}}+\epsilon{I}$.  Then, the $df/dz \geq dg/dz$ becomes 
\begin{align}
{v}^{T}{L}_1(z)^{-1}{\Gamma}_\mathcal{D}{L}_1(z)^{-1}{v}\leq
{v}^{T}{L}_2(z)^{-1}{\Gamma}_\mathcal{D}{L}_2(z)^{-1}{v},
\label{ineq:sub}
\end{align} where we used the fact that for any ${A} \succ 0$, ${B}\succeq 0$, $z \in (0,1)$, $\frac{d}{dz}({A}+z{B})^{-1}$ $=$ $-({A}+z{B})^{-1}{B}({A}+z{B})^{-1}$.

To show that this holds, first observe that both ${L}_1(z)$ and ${L}_2(z)$ are full rank.  Thus, $\rho({\Gamma}_\mathcal{D}^{1/2}{L}_1(z)^{-1})= \rho({\Gamma}_\mathcal{D}^{1/2}{L}_2(z)^{-1})=$  $\rho({\Gamma}_\mathcal{D}^{1/2})$ and, as a result,  $\mathcal{R}({\Gamma}_\mathcal{D}^{1/2}{L}_1(z)^{-1})=\mathcal{R}({\Gamma}_\mathcal{D}^{1/2}{L}_2(z)^{-1})=\mathcal{R}({\Gamma}_\mathcal{D}^{1/2})$~\cite{bernstein2009matrix}.  Hence, if ${v} \notin \mathcal{R}({\Gamma}_\mathcal{D}^{1/2})$, then~\eqref{ineq:sub} holds trivially.  Otherwise, if ${v} \in \mathcal{R}({\Gamma}_\mathcal{D}^{1/2})$, then $\exists \hat{{v}}$ such that ${v}={\Gamma}_\mathcal{D}^{1/2}\hat{{v}}$ and~\eqref{ineq:sub} is written equivalently
\begin{align}
\hat{{v}}^{T}{\Gamma}_\mathcal{D}^{1/2}&{L}_1(z)^{-1}{\Gamma}_\mathcal{D}{L}_1(z)^{-1}{\Gamma}_\mathcal{D}^{1/2}\hat{{v}} \nonumber \leq\\
&\hat{{v}}^{T}{\Gamma}_\mathcal{D}^{1/2}{L}_2(z)^{-1}{\Gamma}_\mathcal{D}{L}_2(z)^{-1}{\Gamma}_\mathcal{D}^{1/2}\hat{{v}}.
\label{ineq:sub_new}
\end{align}

To prove~\eqref{ineq:sub_new}, it is sufficient to show that $\forall z\in [0,1]$
\begin{align}
{\Gamma}_\mathcal{D}^{1/2}{L}_1(z)^{-1}&{\Gamma}_\mathcal{D}{L}_1(z)^{-1}{\Gamma}_\mathcal{D}^{1/2}\nonumber \preceq\\
&{\Gamma}_\mathcal{D}^{1/2}{L}_2(z)^{-1}{\Gamma}_\mathcal{D}{L}_2(z)^{-1}{\Gamma}_\mathcal{D}^{1/2}\label{ineq:sub_new_2}.
\end{align}
To this end, first observe that ${L}_1(z) \succeq {L}_2(z)$.  This implies ${L}_2(z)^{-1} \succeq{L}_1(z)^{-1}$~\cite{bernstein2009matrix} and, as a result,
\begin{align}
{\Gamma}_\mathcal{D}^{1/2}{L}_2(z)^{-1}{\Gamma}_\mathcal{D}^{1/2} \succeq {\Gamma}_\mathcal{D}^{1/2}{L}_1(z)^{-1}{\Gamma}_\mathcal{D}^{1/2}.  \nonumber
 \end{align}

Now, since for any $0 \preceq {A} \preceq {B}$, ${A}^2 \preceq {B}^2$~\cite{bernstein2009matrix}, the previous inequality gives~\eqref{ineq:sub_new_2}.   \hfill{}{\scriptsize $\blacksquare$}{\scriptsize \par}

\subsection{Theorem~\ref{th:minimal}}

We first prove~\eqref{explain:th:minima3},~\eqref{explain:th:minimal1} and~\eqref{explain:approx_error0}, and then~\eqref{explain:th:minimal2}.  First, let $\Delta_0, \Delta_1, \ldots$ be the sequence of sets selected by Algorithm~\ref{alg:minimal-leaders}, and let $l$ be the smallest index such that $\phi(\Delta_l) \leq E$.  Then, $\Delta_l$ is the set that Algorithm~\ref{alg:minimal-leaders} returns, and this proves~\eqref{explain:th:minima3}.

Moreover, from~\cite{citeulike:416650}, since for any $\Delta \in \mathcal{V}$, $h(\Delta)\equiv-\phi(\Delta)+\phi(\emptyset)$ is a non-negative, non-decreasing, submodular function (cf.~Proposition~\ref{prop:subm}), it is guaranteed for Algorithm~\ref{alg:minimal-leaders} that
\begin{align*}
\frac{l}{l^\star}&\leq 1+\log \frac{h(\mathcal{V})-h(\emptyset)}{h(\mathcal{V})-h(\Delta_{l-1})}\\
&=1+\log \frac{n\epsilon^{-1}-\phi(\mathcal{V})}{\phi(\Delta_{l-1})-\phi(\mathcal{V})}.
\end{align*}
Now, $l$ is the first time that $\phi(\Delta_l) \leq E$, so $\phi(\Delta_{l-1}) > E$.  This implies~\eqref{explain:th:minimal1}.  Moreover, observe that $0<\phi(\mathcal{V})$ so that from \eqref{explain:th:minimal1} we get $F \leq 1+\log[n\epsilon^{-1}/(E-\phi(\mathcal{V}))]$, which in turn implies~\eqref{explain:approx_error0}. On the other hand, since $0<\epsilon \leq 1/E$ and $\phi(\Delta_l) \leq E$, Proposition~\ref{prop:suf_contr} is in effect, i.e.~\eqref{explain:th:minimal2} holds true.  \hfill{}{\scriptsize $\blacksquare$}{\scriptsize \par}

\subsection{Theorem~\ref{th:minimal_set_main}}

The first and the third statements follow directly from Theorem~\ref{th:minimal}.   For~\eqref{state:1}, first note that when Algorithm~\ref{alg:minimal-leaders_final} exits the \texttt{while} loop and after the following \texttt{if} statement, ${v}^{T}{\Gamma}_\Delta^{-1}{v}- {v}^{T}({\Gamma}_\Delta+\epsilon{I})^{-1}{v}\leq cE$.  Additionally, ${v}^{T}({\Gamma}_\Delta+\epsilon{I})^{-1}{v}< \phi(\Delta)\leq E$; and as a result,~\eqref{state:1} is implied.  Finally, for~\eqref{state:2}, note that ${v}^{T}{\Gamma}_\Delta^{-1}{v}- {v}^{T}({\Gamma}_\Delta+\epsilon{I})^{-1}{v}\leq cE$ holds true when $\epsilon$ is of the same order as $n({v}^T{q}_M)^2/(c\lambda_m^2E)$.  Then, $\log\epsilon^{-1}=O(\log n +\log 1/(c\lambda_mE)$, since $({v}^T{q}_M)^2\leq 1$, which proves~\eqref{state:2} through~\eqref{explain:approx_error0}. 
 \hfill{}{\scriptsize $\blacksquare$}{\scriptsize \par}

\vspace*{-2pt}
\bibliographystyle{IEEEtran}
\bibliography{newRef}

\end{document}